\documentclass[10pt]{iopart}
\pdfoutput=1
\usepackage{graphicx}
\usepackage{hyperref}

\newcommand{\bra}[1]{\ensuremath{\left\langle #1\right|}}
\newcommand{\ket}[1]{\ensuremath{\left|#1\right\rangle}}

\begin{document}

\title{Creating and observing $N$-partite entanglement with atoms}
\author{M S Everitt$^1$, M L Jones$^2$, B T H Varcoe$^2$ and J A Dunningham$^2$}
\address{$^1$ National Institute of Informatics, 2-1-2 Hitotsubashi, Chiyoda-ku, Tokyo 101-8430, Japan}
\address{$^2$ School of Physics and Astronomy, University of Leeds, Leeds LS2 9JT, United Kingdom}
\ead{everitt@nii.ac.jp}

\date{\today.}
\begin{abstract}
The Mermin inequality provides a criterion for experimentally ruling out local-realistic descriptions of multiparticle systems. A violation of this inequality means that the particles must be entangled, but does not, in general, indicate whether $N$-partite entanglement is present. For this, a stricter bound is required. Here we discuss this bound and use it to propose two different schemes for demonstrating $N$-partite entanglement with atoms. The first scheme involves Bose-Einstein condensates trapped in an optical lattice and the second uses Rydberg atoms in microwave cavities. 
\end{abstract}


\section{Introduction}

Entanglement is a central feature of quantum mechanics and the key resource in a wide range of quantum information processing tasks. Being able to detect and characterize the entanglement present in a system is therefore an important challenge for physicists. The first mathematically sharp method of detecting entanglement for pairs of particles was proposed by John Bell in 1964 \cite{Bell1964a,Bell1966}. His now-famous inequality provided an unambiguous way of 
distinguishing quantum-mechanical predictions from those of local realistic models. 
Multiparticle generalizations of the Bell inequality were subsequently provided by Mermin \cite{Mermin1990} and others \cite{Belinskii1993,Gisin1998,Werner2001}. These can be used to rule out local-realistic models for $N$-particle systems and are also interesting because there is a close relationship between their violation and the security of $N$-partner quantum communications \cite{Scarani2001a}.

In general, a violation of these inequalities means that the particles must be entangled. In fact, by placing a stricter bound on the inequality \cite{Seevinck2002a,Toth2005a,Yu2003a} it is even possible to determine what class of entanglement is present from 2-entangled to $N$-entangled states. This method has been used to experimentally confirm three-body entanglement for photons \cite{Chen2006a}. In this article we discuss the conditions required to demonstrate $N$-partite entanglement and show how it could be generated and detected in atomic systems.
We begin by reviewing the Mermin inequality and then discuss how it could be applied to two different atomic systems. In the first, we consider Bose-Einstein condensates (BECs) trapped in optical lattices where the two states of each atomic qubit are different spatial modes. In the second we consider Rydberg atoms interacting in microwave cavities where the two states are different electronic levels. For both cases we discuss how all the terms in the Mermin inequality could be measured and consider some of the practical issues surrounding their implementation.

\section{Mermin inequality}

It is helpful to start with a brief overview of the Mermin inequality \cite{Mermin1990}. For this, we consider an $N$-particle GHZ state of the form,
\begin{equation}
\ket{\Phi}_{N} = \frac{1}{\sqrt{2}}( |\underbrace{\uparrow  \uparrow\cdots \uparrow}_{N}   \rangle + i|\underbrace{\downarrow  \downarrow\cdots \downarrow}_{N} \rangle), \label{GHZstate}
\end{equation}
where $\uparrow$ or $\downarrow$ in the $j$th position labels the two relevant states of the $j$th particle. Such a state is known to maximally violate the Mermin inequality. In an experiment, these $N$ particles would be spatially separated and measurements made on each of them.  
Mermin noted that the GHZ state, $\ket{\Phi}_{N}$, is an eigenstate of the operator,
\begin{equation}
A_{N} = \frac{1}{2i}\left( \prod_{j=1}^{N} (\sigma_{x}^{j} + i\sigma_{y}^{j})  - \prod_{j=1}^{N} (\sigma_{x}^{j} - i\sigma_{y}^{j}) \right)
\end{equation}
with eigenvalue $2^{N-1}$, 
where $\sigma_{x}^{i}$ and $\sigma_{y}^{i}$ are respectively the  $x$ and $y$ Pauli spin matrices acting on the $j$th particle. 
Expanding, $F=\bra{\Psi} A_{N} \ket{\Psi}$, for some general $N$-particle state \ket{\Psi}, one finds,
\begin{eqnarray}
F &=& \bra{\Psi} \sigma_{y}^1\sigma_x^2\cdots \sigma_x^N \ket{\Psi}+ \cdots \nonumber \\
&-&  \bra{\Psi} \sigma_{y}^1\sigma_{y}^2\sigma_{y}^3\sigma_x^4\cdots \sigma_x^N \ket{\Psi} - \cdots \nonumber \\
&+&  \bra{\Psi} \sigma_{y}^1\cdots\sigma_{y}^5\sigma_x^6\cdots \sigma_x^N \ket{\Psi} +\cdots \nonumber \\
&-& \cdots ,   \label{expansion}
\end{eqnarray}
where each line of (\ref{expansion}) contains all distinct permutations of the operators in the term shown on that line.
The only non-zero terms contain odd numbers of $\sigma_{y}$ operators. For the GHZ state, we have $F ={}_N \bra{\Phi} A_N \ket{\Phi}_N = 2^{N-1}$.

The corresponding value of $F$ for a local hidden-variable state can be found as follows. Following Mermin,
we consider the case where the measured distribution functions, $P_{\mu_{1}\cdots\mu_{N}}(m_{1}\cdots m_{N})$, where $\mu_{j} \in \{x,y\}$ and $m_{j} \in \{\uparrow, \downarrow\}$, that describes the $2^{N-1}$ measurements that must be performed to yield the correlations in (\ref{expansion}), can be written in the conditionally independent form,
\begin{equation}
P_{\mu_{1}\cdots\mu_{N}}(m_{1}\cdots m_{N}) = \int d\lambda \rho(\lambda) \left[ p_{\mu_{1}}^{1}(m_{1},\lambda) \cdots p_{\mu_{N}}^{N}(m_{N},\lambda)\right]  \label{hidden}
\end{equation}
where $p_{\mu_{i}}^{i}(m_{i},\lambda)$ is the probability distribution of results for a measurement in the $\mu_{i}$ basis on particle $i$.
This general hidden-variable form attributes the correlations to some unspecified set of parameters, $\lambda$, common to all $N$ particles, with distribution $\rho(\lambda)$. It accounts for correlations in terms of information jointly available to the particles when they left their common source. 
Mermin showed that such a local hidden-variable state has the bounds \cite{Mermin1990},
\begin{eqnarray}
F &\leq& 2^{N/2}, \hspace*{1cm} \hbox {$N$ even,}\nonumber \\
F &\leq& 2^{(N-1)/2} , \hspace*{5mm} \hbox {$N$ odd.}\label{hiddenineq}
\end{eqnarray}
We see that the GHZ state, $\ket{\Phi}_N$,  for which $F=2^{N-1}$, violates (\ref{hiddenineq}) by an exponential amount.

A violation of  inequality (\ref{hiddenineq}) demonstrates that the particles are entangled, but does not guarantee N-partite entanglement  \cite{Shimizu2005a}. For this, a tighter bound is needed, which can be found as follows.  there is, at most, $(N-1)$-partite entanglement in the system. This means that the state of one of the lattice sites (say the first one) factorizes and the expectation value of the corresponding operator factorizes in each term in (\ref{expansion}). Then using the fact that all expectation values disappear if they contain an even number of $\sigma_y$ operators, it is straightforward to show that we are left with: 
\begin{equation}
F = \langle \sigma_x\rangle F_{N-1} \leq 2^{N-2}, \label{F2}
\end{equation}
where the last step follows because $\langle \sigma_x \rangle \leq 1$ and $F_{N-1}= 2^{N-2}$. 
This agrees with the condition found for three particles in \cite{Toth2005a}. Any state that violates (\ref{F2}) must have at least $N$-partite entanglement.

\section{Implementation with BECs}

The aim of an experiment would be to measure all the expectation values in (\ref{expansion}) and see firstly whether their sum violates the bound given by (\ref{hiddenineq}), in which case entanglement is present, and secondly whether it violates the bound  given by (\ref{F2}) in which case $N$-partite entanglement is present. We now discuss how such a scheme could be implemented with BECs.

The first step is to create a GHZ state of the form of (\ref{GHZstate}). Various proposals have been made for producing such states in the laboratory \cite{Zou2001a, Gerry2001, Lee2002, Jacobson1995}. 
Experiments have successfully created GHZ states  with small numbers of photons \cite{Pan2000a,Mitchell2004a}, and $^9$Be$^+$ ions \cite{Sackett2000a, Roos2004a, Leibfried2005}, and could in principle be scaled up to larger numbers.
Here we consider the case of BECs where states of the form of (\ref{GHZstate}) can be created using beam splitters and nonlinear unitary evolution \cite{Dunningham2001a}. For BECs, the nonlinearity arises naturally as a result of collisions between the atoms.

\begin{figure}[t]
\centerline{\includegraphics{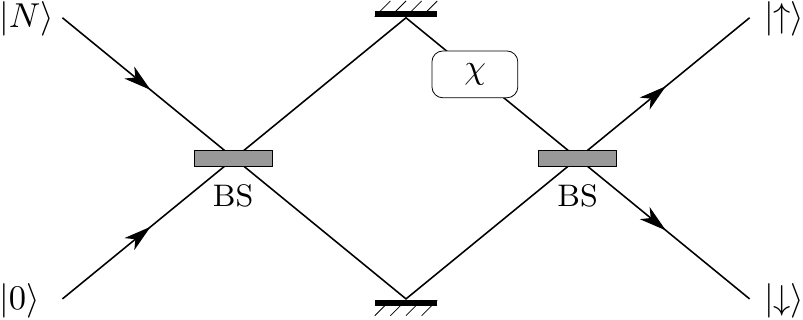}}
\caption{Schematic diagram of the GHZ state creation procedure as a Mach-Zehnder interferometer consisting of two 50:50 beam splitters (BS) and a nonlinearity, $\chi$, on the upper path. If $N$ particles are fed into one input port and the nonlinearity is applied for time, $t= \pi/(2\chi)$, the output is the GHZ state given by (\ref{GHZstate}).} \label{nonlininter}
\end{figure}

A detailed study of the state preparation scheme is presented elsewhere \cite{Dunningham2001a}, but essentially consists of a Mach-Zehnder interferometer with nonlinear evolution between the two beam splitters (see \fref{nonlininter}). The input state consists of $N$ particles at one port and none at the other. If the nonlinearity is applied only to the upper path of the interferometer, the nonlinear part of the Hamiltonian has the form $H_{\rm nl} = \chi \hat{n}_{\uparrow} (\hat{n}_{\uparrow}-1) $, where  $\hat{n}_{\uparrow}$ is the number operator for the upper path and $\chi$ is the strength of the nonlinearity. Evolution due to this Hamiltonian for time $t=\pi/(2\chi)$ then gives state (\ref{GHZstate}) at the output.

Experimentally, this state creation for BECs would involve loading a BEC into one trap of a double-well potential. For now, we consider that the BEC is in a Fock state with $N$ atoms, i.e. \ket{N}, however later we will consider the case of a mixed state. The first beam splitter is implemented simply by rapidly reducing the height of the potential barrier between the two wells, waiting for time $t=\pi/(4J)$, where $J$ is strength of the tunnelling between the wells, and then rapidly raising the barriers again \cite{Dunningham2004}. The nonlinearity can then be `switched on'  for time $t=\pi/(2\chi)$ by the use of Feshbach resonances to change the interaction strength, $\chi$, between the atoms \cite{Cornish2000a}. Finally another beam splitter can be implemented. This gives a state of the form of (\ref{GHZstate}), where the labels $\uparrow$ and $\downarrow$ refer to the two potential wells, shown as the upper and lower outputs from the interferometer in \fref{nonlininter}. 

\begin{figure}[t]
\centerline{\includegraphics{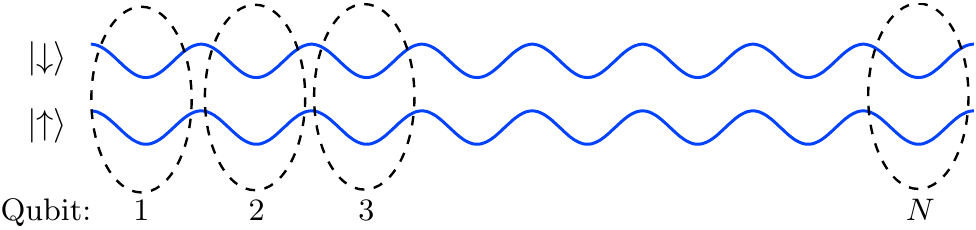}}
\caption{The two spatially separated optical lattices. Each pair of lattice sites (marked by a dashed curve) represents a qubit and contains a single atom. An atom in the upper trap is denoted \ket{\uparrow} and an atom in the lower trap is denoted \ket{\downarrow}.} \label{qubit}
\end{figure}

After the state creation process, we have a superposition of all the atoms in the upper trap and all in the lower trap. The two traps are then illuminated with a retro-reflected laser field that creates a standing wave across them both (see \fref{qubit}). The frequency of this laser is chosen so that the atoms are trapped in the nodes of the standing wave. By adiabatically increasing the intensity of the light, the number fluctuations on each site are progressively squeezed due to the interplay between the interaction and tunnelling energies. Eventually, a Mott insulator transition \cite{Greiner2002a} takes place whereby each pair of upper and lower lattice sites contains precisely one atom. Each qubit in (\ref{GHZstate}) is now spatially distinct and we take $\uparrow$ to represent an atom in the upper lattice site and $\downarrow$ to represent one in the lower lattice site.

The next step is to make measurements on this system that correspond to the terms in (\ref{expansion}). This involves making measurements on each qubit in the basis of the eigenstates of $\sigma_x$ and $\sigma_y$. The eigenstates of $\sigma_x$ are $\ket{x,+}= (\ket{\uparrow} + \ket{\downarrow})/\sqrt{2}$ and 
$\ket{x,-} =(\ket{\uparrow} - \ket{\downarrow})/\sqrt{2}$ with eigenvalues of $+1$ and $-1$ respectively. The eigenstates of $\sigma_y$ are $\ket{y,+}= (\ket{\uparrow} + i\ket{\downarrow})/\sqrt{2}$ and 
$\ket{y,-} =(\ket{\uparrow} - i\ket{\downarrow})/\sqrt{2}$ with eigenvalues of $+1$ and $-1$ respectively.

These measurements can be achieved by passing the two sites of each qubit through a beam splitter. We can see this as follows. The transformation of the single-particle states by a 50:50 beam splitter is,
\begin{eqnarray}
\ket{\uparrow} &\longrightarrow& \frac{1}{\sqrt{2}}\left( \ket{\uparrow} + i\ket{\downarrow} \right) \\
\ket{\downarrow} &\longrightarrow& \frac{1}{\sqrt{2}}\left( i\ket{\uparrow} + \ket{\downarrow} \right). \label{BStransform}
\end{eqnarray}
Using this, it is straightforward to show that if we pass the eigenstates of $\sigma_y$ through a 50:50 beam splitter we obtain, $\ket{y,+} \to \ket{\uparrow}$ and $\ket{y,-} \to \ket{\downarrow}$, where we have ignored any irrelevant global phase. This means that, using a beam splitter and then detecting whether the particle is in the upper or lower site is equivalent to a measurement in the $\sigma_y$ basis. A detection result of \ket{\uparrow} or \ket{\downarrow} gives a measurement outcome in the $\sigma_y$ basis of $+1$ or $-1$ respectively.

For measurements in the $\sigma_x$ basis, we need a combination of a phase shift and a beam splitter.
We can see this by considering the eigenstates of $\sigma_x$. The combined procedure of applying a phase shift of $\pi/2$ to the upper site and then passing the state through a 50:50 beam splitter transforms the states in the following way: $\ket{x,+} \to \ket{\uparrow}$ and $\ket{x,-} \to \ket{\downarrow}$. A detection result of \ket{\uparrow} or \ket{\downarrow} is therefore equivalent to a measurement outcome in the $\sigma_x$ basis of $+1$ or $-1$ respectively.

In practice, to measure one of the terms in (\ref{expansion}), we would imprint a phase shift on all the sites for which we want to make a $\sigma_x$ measurement. The procedure for doing this is well-understood and has been experimentally demonstrated \cite{Denschlag2000a}. It involves illuminating the target lattice sites with pulsed off-resonant laser light. The phase that is imparted is a function of the detuning of the laser, the laser linewidth, the intensity of the light, and the pulse duration. An appropriate choice of these parameters allows a $\pi/2$ phase to be imprinted.
Next we would simultaneously implement a 50:50 beam splitter between each upper site and its corresponding lower site. This can be achieved simply by rapidly lowering the potential barrier between the upper and lower lattices, waiting for some time $t=\pi/(4J)$ and then rapidly raising the barriers again \cite{Dunningham2004}. The measurement outcome depends on the number of atoms in the lower, $N_{\downarrow}$, and upper, $N_{\uparrow}$ traps and is given by: 
\begin{equation}
(-1)^{N_{\downarrow}}(+1)^{N_{\uparrow}} = (-1)^{N_{\downarrow}}. \label{measoutcome}
\end{equation}
So, in fact, we need only measure the number of atoms in the lower traps. To find the corresponding term in (\ref{expansion}), we need an ensemble average of these measurements and the whole procedure then needs to be repeated for all the terms in (\ref{expansion}). This would allow one to experimentally determine a value for $F$ and see whether it violates the bounds given by (\ref{hiddenineq}) and (\ref{F2}).

One possible difficulty with this scheme is that it requires the experimenter to be able to individually address lattice sites. This is difficult because the lattice sites are spaced by $\lambda/2$, where $\lambda$ is the wavelength of the laser light, and so they are too close to easily resolve. There are, however, a number of suggestions for how this problem may be able to be overcome. In one experiment, atoms were loaded into every third lattice site by superimposing a `superlattice' on top of the regular lattice \cite{Peil2003}. Another idea is to change the spacing between lattice sites by controlling the angle at which the laser beams interfere \cite{Li2008a}. In such a setup one could quickly separate the atoms so that they can be addressed individually to imprint phases or make measurements. A recent experiment has also shown how it is possible to resolve single atoms in a Mott insulator state using fluorescence imaging \cite{Sherson2010}

So far, we have considered the case that the total number of atoms, $N$, is fixed and known. However, in practice, the input BEC will be a mixture of number states and so each experimental run will correspond to a different total number of atoms. This means that the expectation values corresponding to each term in the expansion (\ref{expansion}) will be averages over different particle numbers.  We now show how it is possible to extract the relevant data to use in Eq.~(\ref{expansion}) even though the particle number is different in each trial.

\begin{figure}[t]
\centerline{\includegraphics{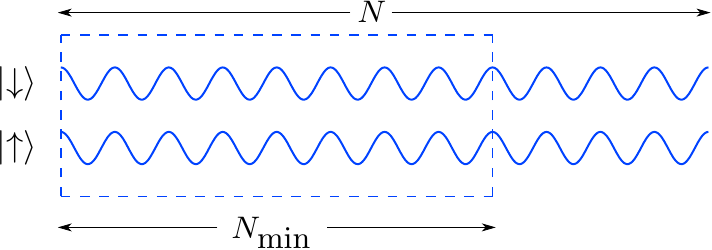}}
\caption{The two optical lattices. In the Mott regime, we assume that for $N$ atoms, each of the first $N$ pairs of upper and lower lattice sites each contains a single atom. To account for fluctuations in the total atom number between experimental runs, we consider only the cases where all $\sigma_y$ measurements take place in the first $N_{\rm min}$ sites, where $N_{\rm min}$ is the minimum total number of atoms on any experimental run.} \label{Nmin_fig}
\end{figure}

We suppose that for $N$ atoms in the final state, a single atom populates each of the first $N$ sites on the optical lattice in the Mott regime \footnote{We take the first $N$ sites for notational simplicity. We could, of course, also consider that the atoms populate the $N$ middle sites of the lattice with a trivial extension to the algebra}. In our scheme, we will consider only the first $N_{\rm min}$ sites, where $N_{\rm min}$ is the minimum number of atoms on any run. Importantly, we retain only experimental runs for which all the $\sigma_y$ measurements take place in the first $N_{\rm min}$ sites (see \fref{Nmin_fig}). This gives us a subset of  $2^{N_{\rm min}-1}$ measurements from all those taken and we neglect the rest. 
In order for this approach to work, we need to show that all the expectation values in (\ref{expansion}) for $N_{\rm min}$ are identical to the results obtained if, for $N>N_{\rm min}$, we measure only a subset consisting of the first $N_{\rm min}$ sites. We can confirm this by calculating the expectation values directly.

For $N$ atoms, the expectation value for a general term in (\ref{expansion}) where all the $\sigma_y$ measurements are in the first $N_{\rm min}$ sites, has the form,
\begin{eqnarray}
{}_N\bra{\Phi} \overbrace{\sigma_{y}^1\cdots \sigma_{y}^j\sigma_{x}^{j+1}\cdots\sigma_x^{N_{\rm min}}}^{\mbox{First $N_{\rm min}$ sites}} \sigma_x^{N_{\rm min}+1}\cdots \sigma_x^N \ket{\Phi}_N  = - i^{j+1}, \label{Nmin1}
\end{eqnarray}
where the last step follows because we only have terms where $j$ is odd. Any permutation of the first $N_{\rm min}$ operators does not change the result.

Similarly, for $N_{\rm min}$ atoms, we get,
\begin{equation}
{}_{N_{\rm min}}\bra{\Phi} \sigma_{y}^1\cdots \sigma_{y}^j\sigma_{x}^{j+1}\cdots\sigma_x^{N_{\rm min}}  \ket{\Phi}_{N_{\rm min}}  = - i^{j+1}. \label{Nmin2}
\end{equation}
Comparing (\ref{Nmin1}) and (\ref{Nmin2}), we see that by making measurements on $\ket{\Phi}_{N}$, we can determine the corresponding expectation values for $\ket{\Phi}_{N_{\rm min}}$. This means that, even when the total number of atoms varies between experimental runs, we can always obtain a complete set of $2^{N_{\rm min}-1}$ terms corresponding to the expansion in (\ref{expansion}) where $N=N_{\rm min}$. In this way, the value of this expression for a GHZ state is,
$F = 2^{N_{\rm min}-1}$,
and the bound for $(N_{\rm min} -1)$-partite entanglement is
\begin{equation}
F \leq 2^{N_{\rm min}-2}. \label{Fmixed2}
\end{equation}

The corresponding bound for the hidden variable model follows easily from (\ref{hidden}) since the joint probability for $N$ sites can be reduced to a joint probability for $N_{\rm min}$ sites simply by integrating over $m_{N_{\rm min}+1}, \cdots, m_{N}$. This gives,
\begin{eqnarray}\fl
P_{\mu_{1}\cdots\mu_{N_{\rm min}}}(m_{1}\cdots m_{N_{\rm min}}) = \int d\lambda \rho(\lambda) \left[ p_{\mu_{1}}^{1}(m_{1},\lambda) \cdots p_{\mu_{N}}^{N}(m_{N_{\rm min}},\lambda)\right],  \label{hidden2}
\end{eqnarray}
i.e. the first $N_{\rm min}$ sites do not depend on the measurement outcomes of the last $N-N_{\rm min}$ sites. The rest of the argument follows that of Mermin \cite{Mermin1990} but with $N$ replaced with $N_{\rm min}$. This gives the hidden variable bounds as,
\begin{eqnarray}
F &\leq& 2^{N_{\rm min}/2}, \hspace*{1cm} \hbox {$N_{\rm min}$ even,}\nonumber \\
F &\leq& 2^{(N_{\rm min}-1)/2} , \hspace*{5mm} \hbox {$N_{\rm min}$ odd.}\label{hiddenineq2}
\end{eqnarray}
Comparing (\ref{hiddenineq2}) and (\ref{Fmixed2}) with the result for a GHZ state, $F=2^{N_{\rm min}-1}$, 
we see that it is still possible to rule out local-realistic models and detect $N_{\rm min}$-partite entanglements even if the total number of atoms in the system is uncertain.

\section{Implementation with cavity QED}

Another possible system for implementing this scheme is Rydberg atoms in microwave cavities.
For this, we consider an $N$-particle GHZ state of the form,
\begin{equation}
\ket{\Phi}_{N} = \frac{1}{\sqrt{2}}\bigl( |\underbrace{g g \cdots g \rule[-3pt]{0pt}{5pt}}_{N}   \rangle + |\underbrace{e  e \cdots e  \rule[-3pt]{0pt}{5pt}}_{N}   \rangle\bigl), \label{GHZstate2}
\end{equation}
where $g$ or $e$ in the $j$th position denotes that the $j$th particle is in the ground or excited state. This state and the subsequent analysis differs slightly from that discussed in the first part of the paper. We choose to do the analysis for this particular state because it is the one most conveniently created by our proposed experimental scheme for Rydberg atoms. 

This GHZ state, $\ket{\Phi}_{N}$, is an eigenstate of the operator,
\begin{equation}
A_{N} = \frac{1}{2}\left( \prod_{j=1}^{N} (\sigma_{x}^{j} + i\sigma_{y}^{j})  + \prod_{j=1}^{N} (\sigma_{x}^{j} - i\sigma_{y}^{j}) \right)
\end{equation}
with eigenvalue $2^{N-1}$.  For a general $N$-particle state \ket{\Psi}, one finds $F=\bra{\Psi} A_{N} \ket{\Psi}$ is given by
\begin{eqnarray}
F = 1 &-&  \bra{\Psi} \sigma_{y}^1\sigma_{y}^2\sigma_{x}^3\sigma_x^4\cdots \sigma_x^N \ket{\Psi} - \cdots \nonumber \\
&+&  \bra{\Psi} \sigma_{y}^1\cdots\sigma_{y}^4\sigma_x^5\cdots \sigma_x^N \ket{\Psi} +\cdots \nonumber \\
&-&  \bra{\Psi} \sigma_{y}^1\cdots\sigma_{y}^6\sigma_x^7\cdots \sigma_x^N \ket{\Psi} -\cdots \nonumber \\
&+& \cdots ,   \label{expansion2}
\end{eqnarray}
For the GHZ state (\ref{GHZstate2}), we have $F ={}_N \bra{\Phi} A_N \ket{\Phi}_N = 2^{N-1}$, and 
any hidden-variable state has the same bounds as before, i.e. 
\begin{eqnarray}
F \leq 2^{N/2}, \hspace*{1cm} \hbox {$N$ even,}\nonumber \\
F \leq 2^{(N-1)/2} , \hspace*{5mm} \hbox {$N$ odd.}\label{hiddenineq2}
\end{eqnarray}

The first step to seeking violations of (\ref{hiddenineq2}) using Rydberg atoms is to create a GHZ of the form of \eref{GHZstate2}. We start with $N$ atoms of ${}^{85}$Rb, each initially in the Rydberg state 63P. Using resonant microwave fields, transitions can be driven to the levels 61D and 62P. We shall refer to 63P as \ket{e}, the relative excited state, and 61D as \ket{g}, the relative ground state. These form the basis used to construct the GHZ state. The state 62P is referred to as \ket{i}, an auxiliary state. The state \ket{e} is produced using a three step laser excitation \cite{Sanguinetti2009}. Zheng and Guo demonstrate in their paper \cite{Zheng2000} that by using three states in this manner with a cavity detuned from the $\ket{e}\leftrightarrow\ket{g}$ transition it is possible to create an EPR state with a pair of atoms. This was later demonstrated experimentally by Osnaghi et al. \cite{Osnaghi2001}.
We extend this scheme to show that by using the same states and $N-1$ cavities in a line, it is possible to produce a GHZ state of N atoms \footnote{Alternatively we could use atoms at different velocities and a single cavity so that one atom is present during the passage of the other atoms, which are only present one at a time, effectively emulating multiple cavities.}. An alternative has been demonstrated by Rauschenbeutel et al. \cite{Rauschenbeutel2000}.

\begin{figure}[b]
\centerline{\includegraphics{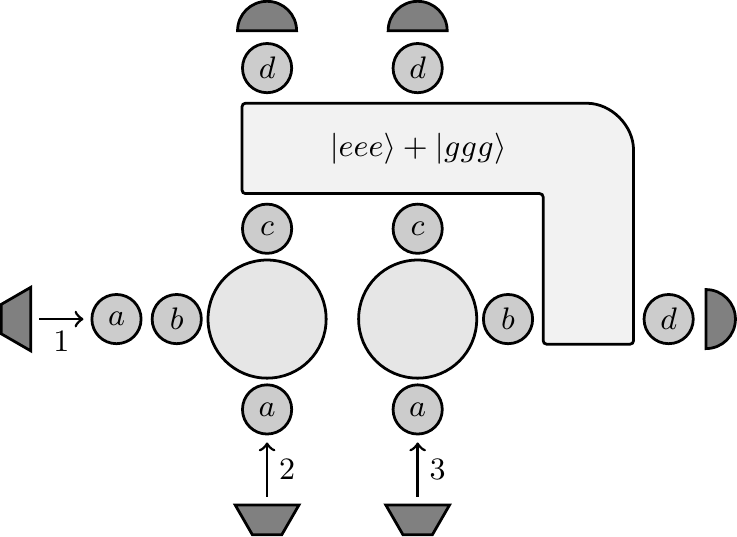}}
\caption[Preparing a five party GHZ state]{This example produces a three party GHZ state of atoms. Atoms approach as indicated by arrows to coincide in each cavity (large circles). Initially all atoms are in state \ket{e}. The rotation zones labelled \textit{a} take \ket{e} to \ket{+} and $c$ take \ket{e} to \ket{-} and \ket{g} to \ket{+}. Those zones labelled \textit{b} act on the first atom (traveling horizontally) to switch \ket{e} components of the state to \ket{i}, the auxiliary state, and vice versa. The enclosed region is where the GHZ state exists. The zones labelled $d$ are arbitrary rotations used to study the state before measurement.\label{GHZ3}}
\end{figure}

We start with all atoms initially in the state \ket{e}. These atoms are produced on demand \cite{Brattke2001}. Next all atoms are rotated to the state $\ket{+}=(\ket{g}+\ket{e})/\sqrt{2}$ using a microwave field resonant with the $\ket{e}\leftrightarrow\ket{g}$ transition, see \fref{GHZ3}. 
This corresponds to a rotation about the $x$-axis on the Bloch sphere. The Hamiltonian used to implement this is
\begin{equation}
H_I=\hbar\Omega\left(\hat{\sigma}^{+}+\hat{\sigma}^{-}\right)\,,
\end{equation}
where $\Omega$ is the coupling strength of the atom with the field and $\hat{\sigma}^{+}$ and $\hat{\sigma}^{-}$ are the atomic raising and lowering operators. The amplitudes of the atom given as $a\,(t)\ket{g}+b\,(t)\ket{e}$ evolve according to the equations
\begin{eqnarray}
a\,(t) &=& a_0\cos(\Omega t) -i b_0\sin(\Omega t)\nonumber\\
b\,(t) &=& b_0\cos(\Omega t) -i a_0\sin(\Omega t)\,,
\end{eqnarray}
which is equivalent to a rotation about the $x$-axis. 
 
The first atom, which will interact with each other atom in turn, then passes through another microwave field (labelled as $b$ in \fref{GHZ3}) which is resonant with the transition $\ket{e}\leftrightarrow\ket{i}$ leaving the first atom in the state $(\ket{g}+\ket{i})/\sqrt{2}$. The first atom now interacts with the second in a high-Q (quality factor) microwave cavity. This cavity is detuned from the $\ket{e}\leftrightarrow\ket{g}$ resonance. The four possible states that may interact are shown in the left column of \tref{theory:table:ZGTruth}. The first two states in the left column of the table are modelled using a two atom Tavis-Cummings model with large detuning. With a zero photon field \cite{Englert1998} this can be solved to give the interaction
\begin{eqnarray}
&&\ket{g,e}\mapsto\mathrm{e}^{-i\gamma t}\left[\cos(\gamma t)\ket{g,e}-i\sin(\gamma t)\ket{e,g}\right]\nonumber\\
&&\ket{g,g}\mapsto\ket{g,g}
\end{eqnarray}
where $\gamma=g^2/\Delta$, $g$ is the atom-field coupling constant and $\Delta$ is the detuning. As the field is detuned from the transition, excited atoms can only virtually excite the field, effectively coupling the atoms together and allowing an excitation to be passed between atoms. The zero photon stipulation is satisfied in the laboratory by cryogenically cooling the cavity, and possibly by preceding the experiment with a chain of atoms in the state \ket{g} to unload the field. The second two states in the left column of \tref{theory:table:ZGTruth} each have one atom in the state \ket{i}, for which all transitions are so far from the cavity resonance that we may assume that it does not contribute to the dynamics of the system. This leaves one atom and the detuned field, which is modelled using the Jaynes-Cummings model. In the case of zero photons and large detuning the evolution of these two states is given by
\begin{eqnarray}
\eqalign
&&\ket{i,e}\mapsto\mathrm{e}^{-i\gamma t}\ket{i,e}\nonumber\\
&&\ket{i,g}\mapsto\ket{i,g}\,,
\end{eqnarray}
For creating a GHZ state we choose the interaction time and detuning such that $t=\pi/4\gamma$. For each interaction this provides the right hand side of the truth \tref{theory:table:ZGTruth}.
\begin{table}[t]
\caption[The Collisional Phase Gate]{The truth table for the collisional phase gate introduced in \cite{Zheng2000}.\label{theory:table:ZGTruth}}
\begin{indented}
\lineup
  \item[]\begin{tabular}{@{}ll}
    \br
    Input & Output\\
    \mr
    \ket{g,g} & \ket{g,g}\\
    \ket{g,e} & \ket{g,e}\\
    \ket{i,g} & \ket{i,g}\\
    \ket{i,e} & -\ket{i,e}\\
    \br
  \end{tabular}
\end{indented}
\end{table}
After the interaction with the first atom each atom emerging from each cavity is rotated so that $\ket{g}\rightarrow\ket{+}$ and $\ket{e}\rightarrow\ket{-}=(\ket{g}-\ket{e})/\sqrt{2}$. After the first atom emerges from the final cavity it enters a microwave field that drives the transition $\ket{i}\leftrightarrow\ket{e}$ so that the atom is left with no amplitude in \ket{i}. The atoms are now in a GHZ state ready for use. \Fref{GHZ3} demonstrates the construction of a three atom GHZ state. It is also possible to generate larger entangled structures with more elaborate cavity arrays \cite{Blythe2006}.

Now we need to make measurements on this system that correspond to the terms in \eref{expansion2}. This involves making measurements on each qubit in the basis of the eigenstates of $\sigma_x$ and $\sigma_y$, i.e. $\ket{g}\pm \ket{e}$ and $\ket{g} \pm i\ket{e}$ respectively. The rotations take place in the regions labelled $d$ in \fref{GHZ3}. The rotation in the basis of $\sigma_x$ has already been discussed, and only one other rotation on the Bloch sphere is needed to realize a rotation in the basis of $\sigma_y$.

The second operation we use is a rotation about the $z$-axis of the Bloch sphere, $\mathrm{R}_z$. This can be implemented simply by applying an electric field to the atom. This effectively increases the transition energy between the levels of the atom and results in a modified phase evolution for the two states. This technique has been experimentally demonstrated with sodium Rydberg atoms by Ryabtsev et al. \cite{Ryabtsev2003}. They applied a resonant microwave pulse to perform an $x$-rotation between two sodium Rydberg atoms followed by a Stark shift $z$-rotation and another $x$-rotation. This combination of interactions allowed them to perform Ramsey interferometry of the $\mathrm{R}_z$ operation applied by the Stark shift. Their results  confirmed that they had successfully implemented the phase operation $\mathrm{R}_z$ and that it was coherent \cite{Ryabtsev2003}.

For Rydberg states of alkali metals the Stark shift varies approximately quadratically with the electric field strength \cite{Davydkin1993},
\begin{equation}
\delta_{g,e} \propto -\alpha_{g,e} E^2\,,
\end{equation}
where $\alpha_{g,e}$ is the polarizability of the ground or excited state of the atom and $E$ is the amplitude of the electric field. The polarizability of the two states will be different, so the relative phase shift will be given by
\begin{equation}
\Theta = \epsilon(\alpha_g-\alpha_e)E^2\,,
\end{equation}
where $\epsilon$ is a factor calculated by integrating over the pulse shape of the electric field. Applying the electric field for some time $t$, the state of the atom evolves as
\begin{equation}
a\ket{e}+ b \ket{g} \mapsto \mathrm{e}^{-i\Theta t} a\ket{e} + b\ket{g}.
\end{equation}
This corresponds to a rotation about the $z$-axis, where $\Theta t$ is the angle of rotation. 

Now, in order to measure an atom in the $\sigma_{x}$ basis, we simply apply the resonant microwave field to implement a rotation of the state by $\pi/2$ about the $y$-axis and then measure the atom to see whether it is in the state $\ket{g}$ or $\ket{e}$. We can see that this works because the rotation maps the eigenstates of $\sigma_x$ directly onto the states $\ket{g}$ and $\ket{e}$.
Similarly, to measure in the $\sigma_y$ basis we apply a $\pi/2$ rotation about the $z$-axis (using the Stark shift), then a $\pi/2$ rotation about the $y$-axis, and then measure the atom to see whether it is in $\ket{g}$ or $\ket{e}$.
Measuring Rydberg atoms is discussed by Gallagher \cite{Gallagher1988}. The particular measurement process that is useful in this experiment, state selective field ionisation, is described in \cite{Englert1998}.

Finally we ought to comment on how noise affects these results since it is well-known that detector inefficiencies in particular give rise to the so-called detection loophole which can undermine our ability to exclude local-realistic descriptions using Bell-type inequalities. Braunstein and Mann \cite{Braunstein1993a} have considered this problem and shown that if the noise is sufficiently small, then the signal for violation grows exponentially faster with $N$ than the noise. In particular, the noise per detector or per particle needs to be less than about 14\%. This bodes well for the feasibility of the detection process. The detector efficiency of field ionization detectors used in micromaser systems is $\sim40\%$ \cite{Englert1998}. Improving both the efficiency of collection of electrons from ionized Rydberg states and the discrimination between Rydberg states prior to ionization is the subject of ongoing research \cite{Jones2009a}, and we are confident that the necessary detector efficiencies are attainable.

\section{Conclusion}

We have proposed two schemes for demonstrating exponential violations of Mermin's inequality in very different atomic systems. Besides their significance in tests of quantum mechanics versus local realism, the schemes we have proposed could be important tools in unambiguously creating and identifying genuine $N$-body entanglement in atomic systems. The experimental techniques required, while challenging, are not far from what can currently be achieved in the laboratory.

\ack
This work was supported by a JSPS Postdoctoral Fellowship and the United Kingdom EPSRC through an Advanced Fellowship GR/T02331/01, Grant GR/S21892/01,
a RCUK Fellowship, and the EuroQUASAR programme EP/G028427/1.

\section*{References}
\providecommand{\newblock}{}

\end{document}